\documentclass[usenatbib,usegraphicx]{mn2e}
\newcommand{\lok}{\left(}
\newcommand{\rok}{\right)}
\newcommand{\lensa}{\left<}
\newcommand{\rensa}{\right>}
\newcommand{\lkw}{\left[}
\newcommand{\rkw}{\right]}
\newcommand{\beq}{\begin{equation}}
\newcommand{\eeq}{\end{equation}}
\newcommand{\beqa}{\begin{eqnarray}}
\newcommand{\eeqa}{\end{eqnarray}}
\newcommand{\kunit}{\, h \, \mathrm{Mpc}^{-1}}
\newcommand{\Mpc}{\, h^{-1} \mathrm{Mpc}}
\newcommand{\Mpcden}{\, h^{3} \mathrm{Mpc}^{-3}}
\newcommand{\Gpcvol}{\, h^{-3} \mathrm{Gpc}^3}
\newcommand{\munit}{h^{-1} M_{\odot}}
\newcommand{\Dsigma}{\Delta_{\sigma}}
\newcommand{\lls}{L/L_\star}

\def\apj{ApJ}

\def\mnras{MNRAS}
\def\nat{Nat}

\def\aap{A\&A}

\def\apjs{ApJS}

\def\prd{Phys.\ Rev.\ D}

\def\physrep{Phys.\ Rep.}

\title[Systematic effects in the sound horizon scale measurements]{Systematic effects in the sound horizon scale measurements}
\author[J.Guzik, G.Bernstein and R.E.Smith]{Jacek Guzik$^{1,2}$\thanks{
E-mail: guzik@astro.upenn.edu (JG), garyb@physics.upenn.edu (GB), res@astro.upenn.edu (RES)}, 
Gary Bernstein$^{1}$ and Robert E. Smith$^{1}$ \\
$^{1}$ Department of Physics and Astronomy, University of Pennsylvania, Philadelphia, PA 19104, U.S.A. \\
$^{2}$ Astronomical Observatory, Jagiellonian University, Orla 171, 30-244 Krak\'ow, Poland}

\begin{document}

\maketitle

\label{firstpage}

\begin{abstract}
We investigate
three potential sources of bias in distance estimations made assuming
that a very simple estimator of the baryon acoustic oscillation (BAO)
scale provides a standard ruler. These are the effects of the
non-linear evolution of structure, scale-dependent bias and errors in
the survey window function estimation. The simple estimator used is
the peak of the smoothed correlation function, which provides a
variance in the BAO scale that is close to optimal, if appropriate
low-pass filtering is applied to the density field. 
While maximum-likelihood estimators can eliminate biases if the form of the systematic 
error is fully modeled, we estimate the potential effects of un- or mis-modelled systematic errors.
Non-linear structure growth using the \citet{2003MNRAS.341.1311S}
prescription biases the acoustic scale by $< 0.3\%$ at $z \ge 1$
under the correlation-function estimator.
The biases due to representative but simplistic models of scale-dependent galaxy bias are
below $1\%$  at $z \ge 1$ for bias behaviour in the realms suggested by halo model
calculations, which is expected to be below statistical errors for a $1000$ sq.~degs. spectroscopic 
survey.  The distance bias due to a survey 
window function errors is given in a simple closed form and 
it is shown it has to be kept below $2\%$ not to bias acoustic scale more than $1\%$ at $z=1$, 
although the actual tolerance can be larger depending upon galaxy bias. 
These biases are comparable to statistical errors for ambitious surveys if no correction
is made for them. 
We show that RMS photometric zero-point errors (at limiting magnitude $25$ mag) below
$0.14$ mag and $0.01$ mag for redshift $z=1$ (red galaxies) and $z=3$ (Lyman-break galaxies),
respectively, are required in order to keep the distance estimator bias below $1\%$.
\end{abstract}

\begin{keywords}
cosmological parameters -- large-scale structure of Universe -- dark matter -- galaxies: statistics
\end{keywords}

\section{Introduction}
\label{sec1}

Acoustic oscillations in the plasma during the
pre-recombination epoch have been detected in the cosmic
microwave background (CMB) \citep{2003ApJS..148..175S}.
Moreover, these oscillations should be imprinted on the distribution
of matter in the Universe and survive
until the present epoch. The presence of these oscillations was
detected as a `hump' in
the correlation function of Luminous Red Galaxies (LRG) sample in the
Sloan Digital Sky Survey (SDSS)
\citep{2005ApJ...633..560E}, with some evidence  
in the power spectrum of galaxy distribution in the 2dF
Galaxy Redshift Survey \citep{2005MNRAS.362..505C}. 
These pioneering measurements
reveal the potential of future galaxy surveys, larger and deeper than
present ones, to measure the sound horizon scale as a function of
redshift for epochs up to the present day.
The physics of the plasma era are well
understood and let us compute the sound horizon in absolute units
(e.g. meters).
BAO measurements may become the best modern
incarnation of standard cosmological tests based on a standard ruler,
for determining the
expansion history of the Universe and/or in tests for curvature
\citep{2006ApJ...637..598B}.

Not much is known, however, about possible systematic effects which
could make the measurement of the sound horizon more difficult than expected. 
The purpose of this paper is to quantify the bias in the sound horizon
scale introduced by effects like nonlinear evolution, scale-dependent
bias in tracer objects distribution and observational selection
(window function).  We compare these levels of potential systematic error to
statistical uncertainties expected for future hemisphere-scale galaxy surveys.
Is the
baryon peak in the correlation function a robust measure of a
distance once set in place by physics in the pre-recombination era?

\begin{figure}
\includegraphics[width=8cm]{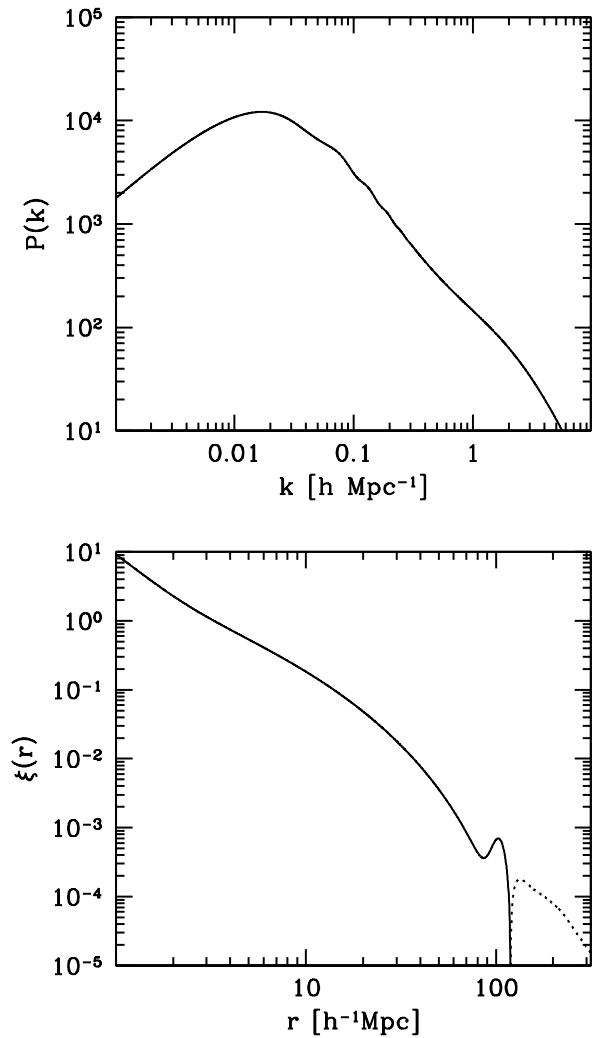}
\caption{
\label{fig1} The nonlinear dark matter power spectrum (upper panel) and
respective correlation function (lower panel) for our fiducial
cosmological model at the redshift $z=1$. Baryon acoustic
oscillations are seen as wiggles in the power spectrum and the `hump'
at scale $102.1 \; \Mpc$ in the correlation function.  For scales
larger than the baryon acoustic peak the correlation function becomes
negative (dashed line).
}
\end{figure}

Quantitative examination of biases in BAO distance estimation require
that we specify the means by which a single distance estimation will
be extracted from the surveyed galaxy distribution in some redshift
range.  The existing literature on the precision of BAO estimates
considers the distance-estimation problem in Fourier space
\citep{2003ApJ...598..720S, 2003ApJ...594..665B,
2005MNRAS.362L..25A, 2006MNRAS.365..255B, 2005ApJ...633..575S}, where
the BAO signature is
a set of ``baryon wiggles'' appearing in the galaxy power spectrum
at scales $ 0.01 \kunit \la k \la 0.5
\kunit$.  A model power spectrum, uncertain by a scale factor, is fit
to the observed galaxy power spectrum.  In the limit where the galaxy
distribution is Gaussian and the model power spectrum is exactly
specified by theory, this yields the maximum-likelihood estimator for
the characteristic scale and hence an unbiased, minimum-variance
estimation.
If effects such as nonlinear growth or galaxy bias are correctly incorporated into the model, 
then the maximum-likelihood techniques still yield optimal unbiased estimators. 
It may remain infeasible, however, to model all effects exactly or even know their 
functional forms, so we must consider the potential systematic biases from mis-modelled
 physical effects, or those that are not modelled at all.  Most current BAO analyses or forecasts incorporate
additional free parameters, typically in the form of ``smooth''
functions of $k$ that add to or multiply the baseline ``wiggly''
model,
in an attempt to mimic any kind of broadband systematic effect
\citep{2003ApJ...598..720S, 2006astro.ph..8632T}.
Marginalization over the nuisance smooth-function parameters
yields the single acoustic-scale estimator.

The Fourier-domain marginalization technique somewhat obscures the
means by which the estimator may end up being biased.
The situation
is much clearer in real space, where the linear-era BAO signature is 
a 3-dimensional Green's function comprising a central peak with a
spherical shell of radius $r_s$, the acoustic horizon scale
\citep{2006astro.ph..4361E}.  This
clearly produces a single peak at $r_s$ in the
real-space correlation function.  We therefore choose to make the
location of this correlation-function peak our estimator
and use this very simple estimator to gauge the magnitude of biases from unmodelled systematic errors.  
We will
demonstrate that this simple estimator is not far from optimal, and
hence it is useful to quantify its biases.  This simple estimator
incorporates no particular model for non-linearity or galaxy biasing;
hence by calculating the degree of bias these effects generate in this
estimator, we have a worst-case estimate.  Future analysis of the
detailed physics will be capable of generating corrections that reduce
these biases.

We choose the correlation-function peak to be as model independent
as possible and avoid the obscuration of fitting a smooth function to the
power spectrum.  The appearance of the baryon peak in the correlation
function at scales $\simeq 100 \; \Mpc$ is difficult to mimic with the
small-scale
effects which can obscure the higher-order wiggles in the
power spectrum.  The power spectrum analysis is
appealing due to the independence of evolution of Fourier modes (at
least in the linear regime), but the correlation-function peak also
has a simple statistical analysis in the Gaussian limit.

As this work was
completed, \citet{2006astro.ph..4361E} published an excellent overview of the relation
between the Fourier- and real-space views of acoustic oscillations in
both the linear and non-linear regimes.  While their work focusses on
developing an analytical and physical understanding of the BAO
non-linearities and biasing, we will take the opposite and less
challenging approach of using toy models and halo models for
non-linear effects, which may not be physically accurate, but yield
bias estimates that are sufficiently quantitative to identify biases
that are most in need of further attention.

The fiducial cosmology is the same
as in \citet{2006MNRAS.365..255B} to facilitate comparison:
flat $\Lambda$CDM with total
matter density $\Omega_m = 0.3$ and baryon fraction $f_b =
15\%$. The present Hubble parameter is taken to be $h=0.7$ in units of
$100 \; \mathrm{km/s/Mpc}$, power spectrum normalization $\sigma_8 =
1$ and the primordial spectral index $n_s = 1$. 
We use the
linear transfer function given by \citet{1998ApJ...496..605E}.  
Nonlinear corrections to the power spectrum are based on the fitting
formula given by \citet{2003MNRAS.341.1311S} or the halo model
\citep{2002PhR...372....1C}.  The nonlinear fitting formula was
inferred from results of the pure dark matter simulations (the
halo model is also usually calibrated with N-body simulations) and it is
not {\em a priori} clear that it is very precise for the case of
nonzero baryon contribution.
The matter power spectrum and respective correlation
function for our fiducial model are shown in Fig. \ref{fig1} where a
`hump' around the scale $100 \Mpc$ is the BAO feature.
Changing cosmological parameters makes the `hump' scale,
amplitude and width change accordingly \citep{2004ApJ...615..573M}.
A fast and reliable transformation between the Fourier and the real space 
can be performed by means of the \texttt{FFTLog} routines \citep{2000MNRAS.312..257H}.

Our analysis is simplified in the sense that we work in real space
without distinguishing between transverse and radial directions and we
do not take redshift space distortions into account. The former assumption 
means that both tangential and radial directions are taken to scale the same with the 
angular-diameter distance $D_A(z)$, so information about $D_A(z)$ might be 
obtained from either direction. 
In other words we assume that accurate spectroscopic redshifts are available for sources
with negligible peculiar velocities, and that the product of the angular-diameter 
distance and the Hubble parameter $D_A(z)H(z)$ is
known rather than having $D_A(z)$ and $H(z)$ measured independently.
These assumptions are unrealistic, but should not spoil the rough
estimates of bias from physical effects other than redshift-space distortion.

\section{Statistical errors on the sound horizon scale}
\label{sec2}

Before addressing the bias in the correlation-function estimator of
$r_s$, we analyse the estimator's statistical
errors due to the finite survey volume $V_s$ (sample
variance) and finite density of observed objects $n_g$ (shot noise).

Errors associated with sample variance scale roughly with the comoving
survey volume as
$V_s^{-1/2}$ \citep{1997PhRvL..79.3806T}.  In the ``concordance''
$\Lambda$CDM model
\citep{2003ApJS..148..175S}, the comoving volume of space 
available for observations which cover the whole sky grows from 
$2.4 \Gpcvol$ for $z \le 0.3$, through $52 \Gpcvol$ for $z \le 1$ up
to $284 \Gpcvol$ for $z \le 2.5$.
For this reason, as well as the degradation of the BAO feature by
nonlinearities at $z<1$, the statistical power of BAO surveys degrades
significantly at $z<1$, although see \citet{2006astro.ph..4361E} for a
potential strategy to ameliorate the second effect.

All of the statistical errors on BAO estimators scale with
$V_s^{-1/2}$. 
For illustrative purposes we will consider volumes contained in
very wide redshift bins, $\Delta z = 0.5$.
Proposals exist for near-term
surveys to cover $\sim 1000$ sq.~degs. of the sky, for which
a redshift bin $\Delta z = 0.5$ centred at $z = 1$ gives an
observable volume $V_s = 1.4 \Gpcvol$, twice as much as used by SDSS
collaboration to discover the baryon acoustic peak.  A highly
ambitious goal would be a spectroscopic BAO survey
covering half of the sky, for a 
survey volume of $28 \Gpcvol$ in the $z=1$ bin.  We will term these
the ``modest-scale'' and ``hemisphere-scale'' survey scenarios.

The shot noise comes from the finite
number of objects sampling the distribution of mass.  With finite
resources (telescope time) a survey of the baryon peak has to be
balanced between volume and sampling density to obtain maximal signal
to noise ratio.  As 
elaborated by \citet{2003ApJ...598..720S} and
\citet{2004ApJ...615..573M} these considerations typically lead to
a relation between the optimal number density of objects $n_g$ and 
their power spectrum $P(k)$ in the form $n_g P(k)\sim 1$.  In the
case of BAO we will assume $n_g P(k=0.2 \kunit) = 3$
\citep{2003ApJ...598..720S}, unless
specified otherwise.  For $z = 1$ we obtain
$n_g = 2.5 \times 10^{-3} \Mpcden$ if galaxies are unbiased relative
to the mass distribution.
In sec. \ref{halomodel} we will consider galaxy bias and its effect on
the baryon acoustic peak.  

In order to compute statistical errors in the sound horizon scale
measurement let us consider the behaviour of the galaxy correlation function $\xi_g$  about
its peak at $r \simeq 100 \Mpc$.  The characteristic scale where the
peak in the correlation function is observed will be denoted $r_c$ and
is an estimator, possibly biased, of the sound
horizon scale $r_s$.  
We allow for smoothing of the observed galaxy distribution with
a window $W_R(k)$ with a characteristic scale $R$ to suppress
the small scale power (i.e. mostly the shot noise) which
contributes to the noise but not to the BAO signal in the correlation
function.  
Next, let us compute the derivative $F(r)$ of
the correlation function as an integral over the power spectrum:
\beq
	F(r) = {d\xi_g \over dr} 
        = - \frac{1}{2\pi^2} \int \; dk k^3 P(k) W^2_R(k) j_1(kr),	
        \label{fr} 
\eeq
where $j_m(x)$ is a spherical Bessel function of the $m$-th order.
The position of the characteristic peak is given by $F(r_c)
= 0$. Thus uncertainty in its position is given by
\beq
	\sigma_{r_c} = \frac{\sigma_F}{|\partial F/\partial r|},
	\label{sig1}
\eeq 
taken at $r=r_c$. From eq. (\ref{fr}) we obtain the variance of $F(r_c)$ as
\beq
	\sigma^2_F = \frac{2}{2\pi^2 V_s} \int \; dk k^4 \sigma^2_P(k) W^4_R(k) j^2_1(k r_c), 
	\label{sig3}
\eeq
where $\sigma^2_P(k)$ is the variance of a single mode of the power
spectrum, given by $\sigma^2_P(k) = \lok P(k) + 1/n_g \rok^2$
\citep{1994ApJ...426...23F, 1997PhRvL..79.3806T}. 
Hence, the statistical uncertainty on the position of the
characteristic scale $r_c$ is given by 
\beq
	\sigma^2_{r_c} = \frac{\sigma^2_F}{
                        \lkw \int dk \; k^4 P(k) W^2_R(k) \lok j_0(k r_c) - \frac{2}{kr_c} j_1(k r_c) \rok \rkw^2},
	\label{sigrs}
\eeq 
where $\sigma^2_F$ is given by eq. (\ref{sig3}). 
We adopt a Gaussian filter of the form $W_R(k) =
e^{-k^2 R^2/2}$, i.e. we presume the galaxy density field to be
smoothed with a Gaussian of width $R$ before measuring the correlation
function. 

\begin{figure}
\includegraphics[width=8cm]{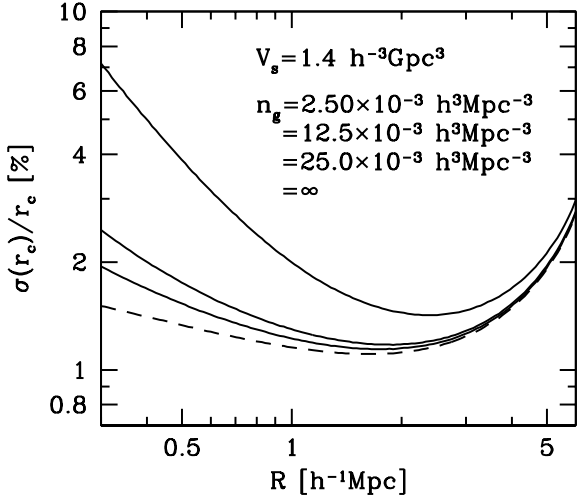}
\caption{
\label{fig2}
Relative statistical error of the characteristic scale $r_c$ in the
real space correlation function vs. filtering scale $R$.  Note
the competing effects of the shot noise subtraction (small $R$) and 
baryon wiggles removal (large $R$), both due to the smoothing.
Shown are
predictions for the modest 1000 sq.~degs. survey covering volume of $1.4
\Gpcvol$ around $z=1$ ($\Delta z = 0.5$) and spatial galaxy density
(solid lines, from top to bottom) $2.5\times 10^{-3} \Mpcden$,
$12.5\times 10^{-3} \Mpcden$, $25\times 10^{-3} \Mpcden$.  The dashed
line is the sample-variance-limited case.  For the full-hemisphere
survey, all values should scaled with survey volume as $V_s^{-1/2}$,
lowering errors by a factor of 4.5. Nonlinear evolution has been
modelled with the \citet{2003MNRAS.341.1311S} prescription.
}
\end{figure}

In Fig. \ref{fig2} we show statistical errors $\sigma(r_c)/r_c$
in the baryon-peak
position, computed by means of eq. (\ref{sigrs}), vs the smoothing
scale $R$, for the modest-scale
survey at redshift $z=1$.  The statistical errors on the hemisphere
survey are expected to be $4.5$ times smaller.

The choice of smoothing scale is a trade between beating down the shot noise
and affecting the BAO signal when smoothing is too
aggressive---smoothing with $R=7 \Mpc$ makes the peak disappear
into a ``knee'' in the correlation function.
A smoothing scale of $R=3 \Mpc$ seems to
yield a signal to noise close to optimal for different shot noise
contributions (see Fig. \ref{fig2}), so we adopt this value for the
remainder of our analysis.

The smoothing procedure shifts the position of the
peak in the correlation function, biasing the recovered
characteristic scale $r_c$. For our assumed $R=3 \;
\Mpc$ filter the bias is $-0.8\%$. 
This will not be a concern in
the course of this work because we are only interested in relative
changes of the peak position with redshift.  Furthermore, 
we could in practice correct the peak shift using
theoretical spectra and knowledge of the smoothing procedure. 

The expected errors for the acoustic scale are similar to those claimed by
\cite{2006MNRAS.365..255B}. In Table \ref{taba} we present a concise comparison  between
expected errors of our the peak-of-the-correlation function approach (PCF) 
and of the power spectrum approach of \cite{2006MNRAS.365..255B} (B2006).
Because \cite{2006MNRAS.365..255B} consider
errors in tangential and radial acoustic scale separately, we add them in inverse quadrature and
these combined errors are shown in Table \ref{taba}. 
The errors are shown for the modest-scale and the hemisphere-scale surveys around the redshift $z=1$ with a depth
$\Delta z =0.5$. Also, two cases of sampling density are considered, one where $n_g = 2.5\times 10^{-3} \Mpcden$
which is implied by the ``optimality'' condition  $n_g P(k=0.2 \kunit) = 3$,
the other with zero shot noise contribution. We notice that the former case equals to almost sample variance limited.
The statistical errors implied  by the PCF method are the same as obtained by B2006 in the case of the modest-scale 
survey and only $3\%$ smaller for the hemisphere-scale one, assuming $n_g = 2.5\times 10^{-3} \Mpcden$. 
One may infer from this comparison that the PCF method of the sound horizon estimation works well
although the estimate of errors is likely too optimistic (see Sec. \ref{sec1}). 

\begin{table}
\caption{Fractional errors on the estimated acoustic scale for different surveys obtained be means of
the peak-of-the-correlation function method (PCF) described in this paper and the power spectrum results of 
\protect\cite{2006MNRAS.365..255B} (B2006). The surveys are assumed to be around the redshift $z=1$ and to have 
the depth $\Delta z =0.5$.  We applied a usual smoothing scale $R=3 \Mpc$.  }
\label{taba}
\begin{tabular}{c||c|c||c|c||}
        & \multicolumn{2}{c||}{$1000 \; \mathrm{sq.~degs.}$} & \multicolumn{2}{c||}{$1/2$ sky} \\
        & \multicolumn{2}{c||}{$V_s = 1.4 \Gpcvol$} & \multicolumn{2}{c||}{$V_s = 29 \Gpcvol$} \\
\cline{2-5}
$n_g [\Mpcden] $  & PCF & B2006 & PCF & B2006 \\ \hline
$2.5\times 10^{-3}$ & $1.5\%$ & $1.5\%$ & $0.32\%$ & $0.31\%$ \\ \hline
$ \infty$            & $1.2\%$ & $1.1\%$ & $0.27\%$ & $0.24\%$ \\ \hline
\end{tabular}
\end{table}

Because the peak-of-correlation-function estimator is nearly as precise
as maximum-likelihood power-spectrum-fitting techniques, it is
worthwhile to use this simple model-independent estimator to
investigate bias properties. 

\section{Effects of nonlinear evolution}
\label{section3}

The nonlinear evolution of matter perturbations in the
Universe and its effect on the baryon wiggles in the power spectrum
have been studied using numerical simulations
(e.g. \citet{1999MNRAS.304..851M, 2005Natur.435..629S,
2005ApJ...633..575S}).  Nonlinear evolution couples
Fourier modes of the matter distribution,
suppresses power at intermediate scales ($k \sim 0.05 \kunit$), 
amplifies power at small scales $k \ga 0.1 \kunit$
and partially erases the baryon wiggles.
This degrades the precision of BAO measurements for given survey
volume, although \citet{2006astro.ph..4361E} suggest that some of this
degradation is reversible using corrections for bulk flows.

\begin{figure}
\includegraphics[width=8cm]{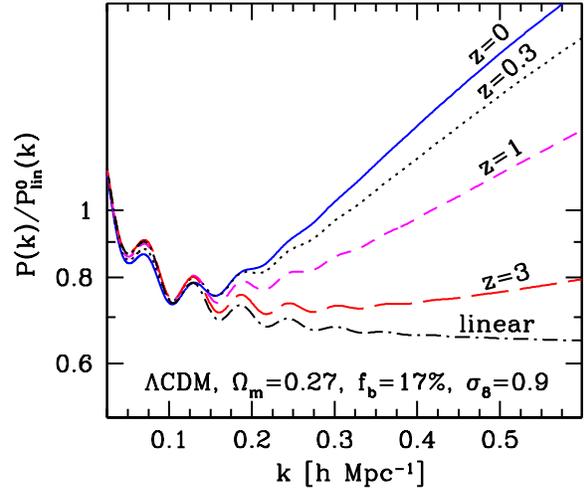}
\caption{
\label{fig4}
The ratio of the nonlinear matter power spectrum for a model with
baryon fraction $f_b = 17\%$, total matter density $\Omega_m = 0.27$
and $\sigma_8=0.9$ to the CMB normalised linear power spectrum
$P^0_{\mathrm{lin}}(k)$ with no baryons for redshifts $z=3$ (long
dashed), $z=1$ (short dashed), $z=0.3$ (dotted), $z=0$ (solid). 
The normalised linear power spectrum is also shown (dot dashed).  
Nonlinear evolution modelling in based on the \citet{2003MNRAS.341.1311S} fitting formula  
and can be compared to the numerical results of 
\citet{2005ApJ...633..575S} in their Fig. 1.  
Note that the cosmological model assumed here differs from the one used throughout
the paper. 
}
\end{figure}

Numerical simulations have not, however, been useful for estimating
{\em biases} in BAO estimators, because they remain too
noisy in the range of scales interesting from the point of view of baryon
wiggles, roughly from $k \sim 0.01 \; \kunit$ through $k \sim 0.5 \;
\kunit$ \citep{2005Natur.435..629S}.  To estimate the bias in a survey
of given volume, the simulation volume must be significantly larger
than the survey volume.  Sufficient resources have not yet been
available to simulate volumes comparable to those to be surveyed by a
hemisphere-scale survey at $z>1$, for example.

One must hence at present rely on analytic arguments to investigate
the bias of BAO distance estimations.
We base our analysis of nonlinear effects on
a fitting formula proposed by \cite{2003MNRAS.341.1311S} which was
inspired by the halo model and calibrated with N-body numerical
simulations. We include the effects associated with baryons in the
formula via the linear transfer function \citep{1998ApJ...496..605E}.
The validity of this description of nonlinear evolution has not yet
been tested extensively, although the formula is able to 
describe reasonably the characteristic features in the power spectra
of simulations that include baryons.  In Fig. \ref{fig4} we
show predictions for the power spectrum behaviour given by the fitting
formula, which can be compared to the numerical results of
\citet{2005ApJ...633..575S}.
Our plot can be compared to their Fig. 1; 
the overall shape of the power spectrum is recovered reasonably well,
however the wiggles, especially those from the third one on, seem to be
more prominent (less erased) in our description than in the
simulations.

\begin{figure}
\includegraphics[width=8cm]{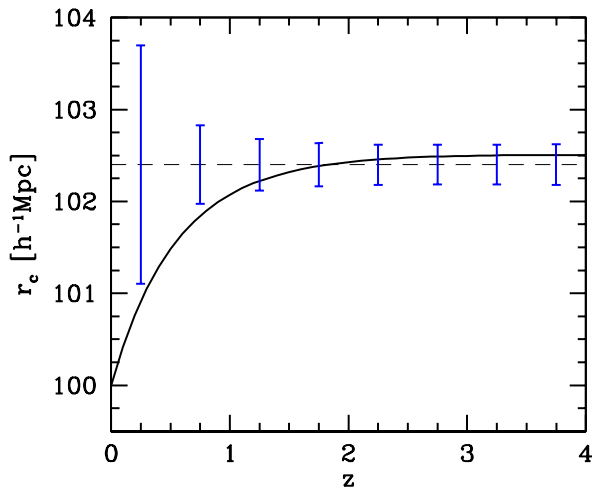}
\caption{
\label{fig3}
Effect of the nonlinear matter evolution on the measured sound horizon
scale from the real space correlation function.  We plot the
characteristic scale position $r_c$ as a function of redshift for our
fiducial cosmological model (solid line).  The horizontal line (dashed)
shows the  characteristic scale for the linear matter evolution, 
$r_c = 102.4 \Mpc$.  The error bars represent statistical
errors in the baryon-peak position from the hemisphere-scale
survey, assuming that we use galaxies lying in bins of width $\Delta z
=0.5$.
}
\end{figure}

The influence of nonlinear effects on the measured
characteristic scale $r_c$ is shown in Fig. \ref{fig3}.
The bias versus the linear-regime value of
$r_c = 102.4 \Mpc$ is negative for small
redshifts and positive for large ones and for sufficiently early
epochs becomes zero.
For epochs earlier than $z = 1.5$, the
bias in the sound horizon scale is $<0.1\%$
and below the statistical error for the hemisphere-scale survey.
At redshift $z=1$ the fractional bias is $\Delta r_c/r_c = - 0.3 \%$,
very close to the expected statistical errors from the hemisphere-scale
survey at that epoch.
For redshifts smaller than $z = 1$
the nonlinear evolution begins to affect first two baryon wiggles,
and the bias approaches $\Delta r_c/r_c = - 2.4 \%$ at $z=0$.  In the
$0<z<1$ range, the bias is comparable to the statistical errors in the
hemisphere-scale survey.  Hence the non-linearity bias would not
greatly dominate the error budget for any feasible survey at any
redshift range, and in fact would be unimportant for any survey that
does not approach hemisphere scale.  This is true even {\em without}
having made any correction to the estimator to account for
nonlinearities. 

It is worth mentioning that the bias due
to the influence  of dark energy on the growth of structure
\citep{2005astro.ph..5565M} is negligible.

\section{The galaxy bias and the baryon-peak position}

The distribution of galaxies is a biased tracer of mass distribution
and depends on the type of galaxies considered
\citep{2005ApJ...630....1Z, 2005A&A...442..801M}.  In order to
quantify the effect of galaxy bias on the measured sound horizon scale,
we must specify the scale dependence of the bias which,
unfortunately, is poorly known at present.  One commonly describes the
galaxy bias in terms of the multiplicative factor $b(k)$ relating
matter power spectrum $P_m(k)$ and the galaxy one $P_g(k)$ as 
$P_g(k) = b^2(k) P_m(k)$.  Obviously any bias that is independent of
scale leaves the location of the
peak in the correlation function unchanged.

\subsection{The galaxy bias in the halo model}
\label{halomodel}

The halo model is a statistical description of the
structure of the Universe based on two basic observations: on large
scales the matter evolution can be described by perturbative models
and on small scales matter clusters into bound haloes of a given
profile (see \citet{2002PhR...372....1C} for a review).  The matter and galaxy
2-point functions are divided into two-halo and one-halo terms, so
the matter and galaxy power spectra become
\beqa
	P_{m,g}(k) &=& P_{\mathrm{lin}}(k) 
	\lok \int dM b_h(M) n_h(M) w^{(2)}_{m,g} u_{m,g}(k) \rok^2 \nonumber \\ 
               && 
        +\int dM n_h(M) w^{(1)}_{m,g} u^2_{m,g}(k),
	\label{pshalo}
\eeqa
where $b_h$ is the halo bias, $n_h$ is the halo mass function, $u_{m}(k)$ ($u_{g}(k)$) is  
the normalised mass (galaxy) distribution profile of a halo of mass $M$ in the Fourier space.
The weight functions for matter and galaxy power spectra have the following 
form for the two-halo term $w^{(2)}_{m}=M/\bar{\rho}$, $w^{(2)}_{g}=\lensa N\rensa/\bar{n}_g$ and 
for the one-halo term $w^{(1)}_{m}=M^2/\bar{\rho}^2$, $w^{(1)}_{g}= \lensa N(N-1)\rensa/\bar{n}_g^2$, respectively. 
$N$ is the number of galaxies of a given type occupying a halo of mass $M$. 
A form of first and second moments of distribution of $N$ must be specified 
and depends on the means of populating haloes with galaxies. 

The relation between the mean number of galaxies $\lensa N \rensa$ in a given halo 
and the halo mass $M$ is dubbed the halo occupation
distribution (HOD).  Usually, the HOD for luminosity-threshold samples 
is described by:  the minimum mass
$M_{\mathrm{min}}$ of a halo in which galaxies may form;
the mass $M_{1}$ of a halo which contain one satellite galaxy
on average;  and the slope $\alpha$ of the relation between
number of satellite galaxies per halo of a given mass:
$ \lensa N(M) \rensa = \lok 1 + \lok M/M_1 \rok^{\alpha} \rok H(M - M_{\mathrm{min}})$,
where $H$ is a step function. 
The details of the halo model parameterisation we use can be found in \citet{2004PhRvD..70d3009H}.
In this framework we can obtain the galaxy bias as a function of scale and
galaxy type provided that we know the HOD. 
The specific form of an HOD was first suggested by theoretical studies
\citep{2000MNRAS.318..203S, 2002ApJ...575..587B, 2004ApJ...609...35K,
2005ApJ...633..791Z}.    

Given the galaxy bias, $b(k)=\sqrt{P_g(k)/P_m(k)}$, or equivalently the galaxy spectrum
$P_g(k)$, from the halo model, we can calculate the correlation
function and hence the characteristic scale $r_c$ of its peak.
In this halo-model calculation, one finds that non-linear growth and
biasing \emph{does not bias the baryon-peak position $r_c$ at
all} (within numerical errors which are about $0.1\%$ in $r_c$) for
halo mass thresholds up to $\sim 100 M_{\star}$. This makes sense
when one realizes that in the halo model the acoustic
oscillations are present only in the two-halo term, which is
assumed to be proportional to the linear matter power spectrum, eq. (\ref{pshalo}).  
The one-halo term is completely featureless at the acoustic scale unless
one considers extremely massive haloes, so the peak of the
correlation function is unchanged.  Another way to state this is that
the halo-model $b^2(k)$ has small oscillations, coherent with the
baryon oscillations, that arrange to leave the correlation peak unchanged.
The scale dependent galaxy bias from the halo model 
was considered by \cite{2006APh....25..172S}. Thorough analysis of the halo bias, 
based on the perturbation theory and numerical simulations, was presented recently 
in \cite{2006astro.ph..9547S}. Their approach extends the standard  halo model by accounting for 
higher order corrections to the linear power spectrum and 
relaxing a common assumption of the linear, scale independent bias
in the two-halo term. They also discuss potential impact of these improvements
on the BAOs.

The near-exact invariance of the correlation function peak may be
considered a peculiarity of the halo model.  
More sophisticated halo models \citep{2001MNRAS.325.1288S, 2003MNRAS.339.1057Y,
2004ApJ...610...61Z, 2005ApJ...631...41T} account for the effect of previrialisation in the 
power spectrum which the standard model of eq. (\ref{pshalo}) fails to describe properly.  
It is usually done by replacing the linear matter power spectrum  with the nonlinear one in eq. (\ref{pshalo})
and accounting for halo exclusion in the two-halo term. 
We do not introduce these improvements in our present analysis because 
the resultant shifts in the PCF are those we ascribed to non-linear growth in Sec. \ref{section3}, 
e.g. Fig. \ref{fig3}. In this section we wish to isolate PCF shifts due solely to 
scale-dependent bias, so we opt to describe the nonlinear bias by a generic smooth function,
(see Sec. \ref{toymodel})
that does not share the halo model's exceptional behaviour, and is simply
described by the maximum and minimum bias values and the $k$ range
over which the bias varies.  We will use the halo model to determine
reasonable ranges for these generic parameters, and then estimate the
shift of the correlation-function peak for the reasonable generic
functions. 

A three-parameter family of HODs allows one to model accurately the
projected correlation function of low-redshift
galaxies ($z \la 0.07$) from
the Sloan Digital Sky Survey \citep{2005ApJ...630....1Z}.
These authors obtain a good fit to the projected
correlation function when $M_{1}/M_{\mathrm{min}} \simeq 23$, almost independent
of the luminosity threshold of galaxies in a sample. At the same time $\alpha$
rises from $0.9$ to $1.2$ with luminosity.  The modelling of the SDSS
galaxy correlation function is done assuming that the distribution of
galaxies within dark matter haloes follow the distribution of the
matter.

Unfortunately, such a detailed modelling of an HOD is not available
for more distant galaxies so we have to base our considerations for
redshift $z \ga 0.1$ on results obtained from numerical
simulations.  Hydrodynamical
simulations carried by \cite{2005ApJ...633..791Z} broadly support the
picture which emerges from already mentioned the SDSS galaxy
clustering data \citep{2005ApJ...630....1Z}. The former show that both
characteristic masses, $M_{\mathrm{min}}$ and $M_{1}$, scale similarly with
galaxy baryonic mass and their ratio for $z=0$ is $M_{1}/M_{\mathrm{min}} \simeq
14$.  Although the numeric value of the ratio differs from the one
obtained from the SDSS galaxy sample, the scaling of $M_{\mathrm{min}}$ and
$M_{1}$ with galaxy baryonic mass (simulations) and luminosity
(observations) is very similar.  On the other hand, dark-matter-only
simulations conducted by \cite{2004ApJ...609...35K} show that the
relation between number of subhaloes and the mass of the host halo is
almost linear ($\alpha=1$) in a very wide range of halo masses
($M_{1}/M_{\mathrm{min}} \ga 3$) and for cosmic epochs $z \la 5$. Another
important result of these simulations is quantifying an evolution of
the ratio $M_{1}/M_{\mathrm{min}}$ with redshift that implies $M_{1}/M_{\mathrm{min}} \simeq
29$ for $z=0$, $M_{1}/M_{\mathrm{min}} \simeq 19$ for $z=1$ and $M_{1}/M_{\mathrm{min}}
\simeq 17$ for $z=3$ (see Fig. 5 in \cite{2004ApJ...609...35K}).

\begin{figure}
\includegraphics[width=8cm]{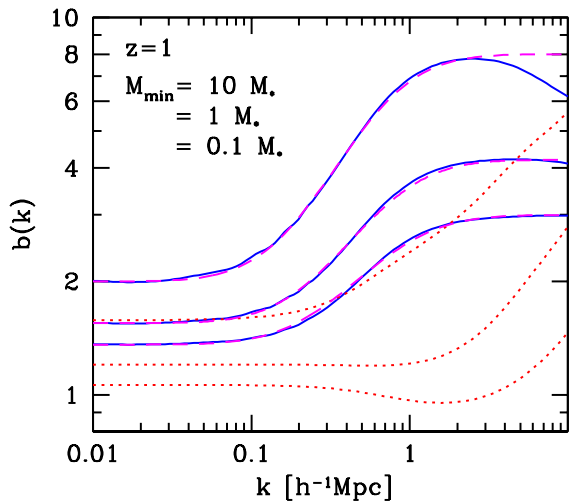}
\caption{
\label{fig5}
The galaxy bias from the halo model for two schemes of populating
haloes with galaxies:  when number of galaxies occupying a halo is
proportional to the halo mass (solid lines), and when there are central
and satellites galaxies with $M_1 = 20 M_{\mathrm{min}}$ (dotted lines).  The minimum
mass of the haloes inhabited by galaxies for each occupation scheme are
$0.1 M_{\star}$, $M_{\star}$ and $10 M_{\star}$ (from bottom to top
for each family of curves). Redshift $z=1$ is assumed for which
$M_{\star} = 8 \times 10^{11} \munit$.
The bias relation as given by the generic 
function described in Sec. \ref{toymodel} is overplotted (dashed lines).
}
\end{figure}

Encouraged by the success of the halo model in describing the galaxy
clustering we use it for the description of the galaxy bias.  In this
work we consider two HOD models, one parameterised by $M_{\mathrm{min}}$, $M_{1}$
and $\alpha$ as mentioned above (``central+satellite''
scheme) and the other where number of galaxies in a given halo is
proportional to its mass for haloes more massive than $M_{\mathrm{min}}$ 
(similar to the ``mass weighted'' scheme of
\cite{2005ApJ...633..575S}).  The main differences between those two
schemes are the following. In the ``central+satellite'' description if
there is at least one galaxy in a halo then exactly one galaxy is at
the halo centre whereas in ``mass weighted'' scheme all galaxies are
distributed within a halo. This implies that in the former case the
small scale power is boosted by the presence of the galaxy in the
centre, in the latter case small scale power is ``smoothed'' in a way
similar to the dark matter distribution.  These effects are shown in
Fig. \ref{fig5} where we present the galaxy bias at $z=1$ as a
function of scale assuming those two schemes of populating haloes with
galaxies. The threshold mass $M_{\mathrm{min}}$ is chosen to be $0.1 M_{\star}$,
$M_{\star}$ or $10 M_{\star}$ where $M_{\star} = 8 \times 10^{11}
\munit$.  We note that the bias in the case of ``mass weighted'' scheme
shows a rapid change on scales which are potentially relevant for a
behaviour of the baryon wiggles ($k$ around a few tenths of $\kunit$).
This is because the most rapid change in the bias appears
where the galaxy power spectrum is becoming dominated by the one-halo term,
whereas the matter power spectrum is
steep (slope $\sim -1.5$) in this regime.
These circumstances appear at different scales, 
depending on the parameterisation of the HOD and the galaxy
distribution, as seen in Fig. \ref{fig5}. 

\subsection{A toy model}
\label{toymodel}

The halo model confines us to the very specific outcome of zero systematic 
which the Universe may not exhibit. We should therefore investigate 
what possible form of the galaxy
bias would imply substantial change in the baryon-peak position.
As mentioned in Sec. \ref{halomodel} the scale dependence of the
galaxy bias in the ``mass weighted'' case as shown in Fig. \ref{fig5}
can be described in the interesting range of scales ($0.01 \kunit
\la k \la 1 \kunit$) as a smoothed step-like function.  A
good description can be provided by the function of the following form
\beq
	\log b^2(k) = \left| \log\frac{b_+}{b_-} \right| \; \mathrm{erf} \lkw \frac{\log k/k_m}{\sqrt{2}\Delta_{\sigma}} \rkw + 
        \log \lok b_+ b_- \rok,
        \label{erfbias}
\eeq
where we use the error function and characterise the bias $b(k)$ by
its asymptotic values at large and small scales, $b_-$ and $b_+$,
respectively (when $b_+/b_- > 1$ we have bias, in the other case --
antibias).  In our analysis it is the ratio $b_+/b_-$ which is
important as multiplying the power spectrum by a constant factor does
not change the position of the peak but its amplitude only.  The other
parameters are the scale $k_m$ where the most rapid change of the bias
occurs, and the logarithmic interval in $k$, $\Dsigma$, centred on
$k_m$, over
which the bias changes by $68.3\%$.  A simple function should suffice
if we are trying to recover trends in
the baryon-peak bias due to the galaxy bias and not to obtain its
precise values.  We exclude situations where the bias has a functional
form containing oscillations of any kind (which was the case in the
halo model description, see Fig. \ref{fig5}).  We expect that the most
prominent effect on the baryon-peak position occurs when the bias
changes rapidly on scales where baryon wiggles are present in the
power spectrum, $ 0.01 \; \kunit \la k \la 0.5 \; \kunit$.

\begin{figure}
\includegraphics[width=8cm]{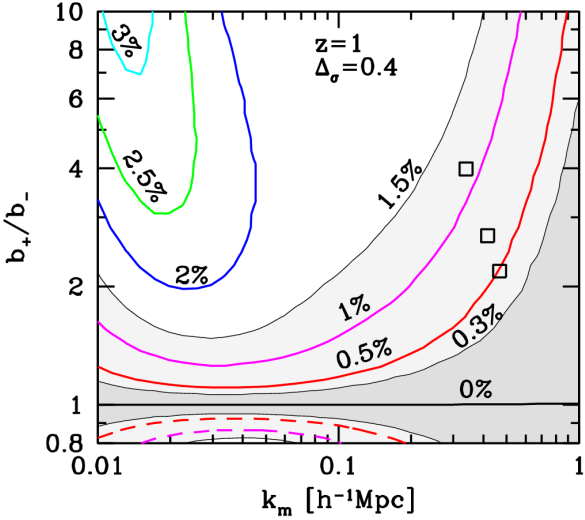}
\caption{
\label{fig6}
Contour plot of the fractional bias of the baryon-peak position as a
function of scale $k_m$ at which the most rapid change of the galaxy
bias appears and the relative amplitude of the galaxy bias
$b_+/b_-$. The range of scales where bias changes the most rapidly is
fixed to $\Delta_{\sigma} = 0.4$.  Open squares mark values of
parameters which describe by means of the smooth function the galaxy
bias obtained from the halo model (see Fig. \ref{fig5}) assuming
minimal halo masses of $0.1 M_{\star}$, $M_{\star}$, $10 M_{\star}$
(from bottom to top).  Shaded regions show parts of the parameter
space where statistical errors on baryon-peak position dominate over
the systematic shift. Statistical errors are assumed to be for two
surveys: the modest-scale (light shaded region) and the hemisphere-scale (dark
shaded region).
}
\end{figure}

We now check whether the baryon-peak of the 
correlation function is a robust measure of the sound horizon in the
presence of scale-dependent, smooth deviations in the power
spectrum.  In Fig. \ref{fig6} we present results of applying the
galaxy bias of a form given by eq. (\ref{erfbias}) to the linear matter power
spectrum.  The typical scale of the bias change is fixed to
$\Dsigma=0.4$, as suggested by the halo model. We take into account only the ``mass weighted'' case as the  
``central+satellite'' scheme introduces negligible bias to the acoustic scale.

The largest effect can be introduced by the galaxy bias
when the most rapid change in bias takes place
around the first baryon wiggle in the power spectrum ($k_m \simeq 0.02
\kunit$) which itself overlaps the matter power spectrum
turnover. The correlation-function peak shift persists 
to $k \simeq 0.08 \kunit$, around the
second peak.  It is difficult to imagine that the
bias has a very rapid change on scales as large as the matter power
spectrum turnover, which is roughly equal to the largest scale of
possible physical interactions in the pre-recombination Universe.
For scales $k_m \simeq 0.5 \kunit$,
where it is more reasonable to consider scale dependence of the
galaxy bias, a very strong bias amplitude of order
several is required to change the baryon-peak position by more than
$2\%$. If we look to the halo model for an estimate of the amplitude
of scale-dependent bias, we find the shift in the characteristic scale
$r_c$ is below 1\% at $z=1$ for any threshold mass below 
$3 M_\star$ (the survey depth is $\Delta z = 0.5$). 
The halo-model parameters for our toy model further
suggest that, for surveys of galaxies inhabitating haloes of mass $M_\star$
a bias of the acoustic peak is $0.7\%$ at the redshift $z=1$, whereas expected statistical error for the modest-scale 
survey is $1.5\%$. For redshifts $z=2$ and $z=3$ respective 
systematic errors are $0.5\%$ and $0.2\%$ and statistical ones $1\%$ and $0.95\%$. 
Statistical errors for the hemisphere-scale survey are smaller by a factor of $4.5$.
The bias becomes comparable to statistical errors for the hemisphere-scale survey, but 
not for the modest-scale one. This is without any correction to the estimator $r_c$. 
Numerical models should eventually allow for corrections to reduce the bias below 
statistical errors ever for hemisphere-scale surveys.   

It is difficult to assign the minimum halo mass that will correspond to future BAO galaxy surveys at $z \ga 1$.
It will depend on the selection criteria for galaxies in the future surveys. 
Haloes of mass $\ga M_\star$ are possible hosts of SDSS Luminous Red Galaxies at $z \sim 0.3$ \citep{2005ApJ...630....1Z}. 
Similar studies of halo occupancy have yet to be done at higher $z$, but the appropriate halo-model mass
threshold $M_{\mathrm{min}}$ will likely be in the range $0.1-10 M_\star$. 

When estimating statistical errors we assumed no galaxy bias. In practice, the 
bias is likely to help in lowering statistical errors either through allowing for a denser sampling 
or covering larger area while using the same resources. 

\section{Survey window function}

Another effect which requires attention is the possible influence of
the survey selection function on the baryon-peak position in the
correlation function. Let us assume that the observers of our
fictitious real-space survey have estimated a spatial selection
function $W_0(\vec{r})$, but the correct selection function differs by a
small multiplicative correction
$\delta_W(\vec{r})$.
The observed galaxy density will be
\beq
\rho_g(\vec{r}) = n_g W_0(\vec{r}) \lok 1 + \delta_W(\vec{r}) \rok \lok 1 +
\delta_{\mathrm{true}}(\vec{r}) \rok. 
\eeq 
The experimenters will estimate the overdensity using their
estimated selection function, so will calculate an (incorrect) local
overdensity 
\beq
\lok 1 + \delta_{\mathrm{obs}}(\vec{r}) \rok
 = 
\lok 1 + \delta_W(\vec{r}) \rok \lok 1 +
\delta_{\mathrm{true}}(\vec{r}) \rok. 
\label{window1}
\eeq
The observers' calculated correlation function $\xi_{\rm obs}$ follows:
\beq
	1 + \xi_{\mathrm{obs}}(r) = 
	\lok 1 + \xi_W(r) \rok \lok 1 + \xi_{\mathrm{true}}(r) \rok, 
\eeq
where we assume that the selection
function is uncorrelated with the true density field, 
and $\xi_W$ is the autocorrelation of
the fractional errors $\delta_W$ in the selection-function estimate.

What level of residuals in the window function are allowed in
order not to bias the sound horizon scale determination more than
predicted statistical errors?
We are interested in scales which are close to the baryon peak in
the correlation function, so we can approximate
correlation function $\xi_{\mathrm{true}}(r)$ to second order
about the true acoustic scale
\beq
	\xi_{\mathrm{true}}(r) = \xi_{\mathrm{true}}(r_s) \lok 1 -
        \frac{\lok r - r_s \rok^2}{2\sigma^2_0} \rok.
\label{truecf}
\eeq
$\sigma^2_0$ has an interpretation as the width of the baryon peak.  The
characteristic scale $r_c$ is measured from the position of the peak
in the observed function $\xi_{\mathrm{obs}}(r)$, so the derivative
of the $\xi_{\mathrm{obs}}$ is zero at $r_c$.  Hence the
shift in the baryon-peak position due to
the presence of spatially-varying error in the selection
function is
\beq
	\delta r_s \equiv r_c-r_s = - \lkw \ln \lok 1 + \xi_{W}(r_c) \rok \rkw^{\prime}
        \frac{\lok 1 + \xi_{\mathrm{true}}(r_s) \rok}{\xi^{\prime \prime}_{\mathrm{true}}(r_s)}, 
 \label{dr1}
\eeq
where the prime means the derivative with respect to $r$ and only
linear terms in $\delta r_s$ are retained in the expression for $\xi_{\mathrm{true}}$.  
Note the bias depends only upon the properties of $\xi_W$ at the
acoustic scale.
 
From eqs. (\ref{truecf}) and (\ref{dr1}) follows that for the power law $\xi_{W}(r)$ with
an exponent $-\alpha$ and an amplitude $\xi_{W} \ll 1$ in the vicinity of the observed peak we obtain
to the leading order 
\beq
	\frac{\delta r_s}{r_s} = - \alpha \lok \frac{\sigma_0}{r_s} \rok^2 \frac{\xi_{W}(r_c)}{\xi_{\mathrm{true}}(r_s)}.
  	\label{rserror}
\eeq
The narrower the acoustic peak, the less easily it is moved; and the
shallower $\xi_{W}(r)$ is at the acoustic scale, 
the smaller a bias in the peak position is produced.

We can make an estimate of the level of systematic errors due to the
unknown properties of the window function for a given survey. As an
example let us consider the survey at the redshift $z \simeq 1$ in which
case we had $r_s = 102.1 \Mpc$, $\xi_{\mathrm{true}}(r_s) = 6.33
\times10^{-4}$ and $\sigma_0 = 10.2 \Mpc$ where we assume no galaxy bias
(Sec. \ref{sec2}) and $\alpha=2$.  Thus eq. (\ref{rserror}) implies that to be able
to beat down the systematic error to the level of $\delta r_s/r_s$,
e.g. $1 \%$, the correlation function of the window function
correction cannot be larger than
\beqa
	\left| \xi_{W}(r_c) \right|&=&3.2 \times 10^{-4} 
	\lok \frac{\xi_{\mathrm{true}}(r_s)}{6.33 \times 10^{-4}} \rok \lok \frac{\sigma_0}{10.2 \Mpc} \rok^{-2} \nonumber\\
	& & \times \lok \frac{\alpha}{2} \rok^{-1} \lok \frac{r_s}{102.1 \Mpc} \rok^{2} \left| \frac{\delta r_s/r_s}{1 \%} \right|.  
	\label{xiwerror}
\eeqa
Hence, eq. (\ref{xiwerror}) shows that unknown variation in the
selection function on scales of the baryon peak should be $\la 1.8\%$ if we
want to bias our results no more than $1\%$. For larger redshifts the allowed 
variation would have to be smaller --- $1.3\%$ and $1\%$ for $z=2$ and $3$, respectively. 
Scaling of the variation with redshift is mainly due to the linear suppression of the mass fluctuations 
and narrowing of the baryon peak (several times smaller effect).  
A survey of biased galaxies makes
$\xi_{\mathrm{true}}(r_s)$ rise, allowing for a larger error in the
selection function without biasing $r_c$ by more than the statistical
error. 

An example of an effect which can bring about correlations on large scales considered above is the
extinction pattern on the sky which is known to be correlated on scales of tens of degrees 
\citep{1998ApJ...500..525S}.  Also, for 
surveys around redshift $z=1(3)$ the acoustic scale is about $2.5(1.3)$ deg 
(as opposed to $\sim 6$ deg for the SDSS LRG sample). This is around (or less) than 
a size of a field of view of planned wide-field surveys, thus a different magnitude limit 
for each telescope pointing may introduce correlations on the interesting scale \citep{2005PhRvD..72d3503G}. 
Moreover, significant uncertainty to a measured window function
may be introduced by a complicated geometry of a survey \citep{2002ApJ...579..483M, 2005MNRAS.362..505C}.  

We investigate a model in which each telescope pointing has an
independent magnitude calibration error, and relate the RMS magnitude
error to the systematic bias on the baryon peak position through the
survey window function as assumed in eq. (\ref{window1}). The
correlation function of the angular window function for calibration
errors in circular fields of diameter
$\theta_a$ is \citep{2005PhRvD..72d3503G}
$\xi_W(\theta) = 2/\pi \Sigma^2 (\arccos (\theta/\theta_a) - (\theta/\theta_a) \sqrt{1-(\theta/\theta_a)^2})$ if $\theta \leq 
\theta_a$ and zero otherwise.
Now let us compute the variance, $\Sigma^2$, which is given as a fractional uncertainty in the density of objects in the
real space due to the survey limiting magnitude error 
\beq
	\sqrt{\Sigma^2} = \frac{\delta (n_g W_0)}{n_g W_0} = \frac{\phi(\lls) 
	\frac{dL}{dm}}{\int_{\lls}^{\infty} \phi(\lls) d\lls} |_{m_{\mathrm{lim}}} \delta m, 
\eeq
where $\phi(L)$ is the luminosity function in the Schechter form $\phi(L) dL =
\phi_{\star} (\lls)^{\alpha} \exp(-\lls) d(\lls)$  and $m_{\mathrm{lim}}$ is a survey limiting magnitude.  
For a survey of red (blue) galaxies at $z \simeq 1.1$ \citep{2005astro.ph..6044F} the luminosity function is 
described by a characteristic density of $\phi_{\star} = 1.5(8.4) \times 10^{-3}(\Mpc)^{-3}$, 
characteristic absolute magnitude $M_{\star} = -21.58(-21.25)$
and the faint-end slope $\alpha = -0.5(-1.3)$. At $z \simeq 3$, observations of Lyman-break galaxies \citep{1999ApJ...519....1S}
yield $\phi_{\star} = 48.0 \times 10^{-3}(\Mpc)^{-3}$, $M_{\star} = -21.68$, $\alpha = -1.6$.
Also, let us assume that the survey telescope aperture 
is $\theta_a = 3$ deg and the limiting apparent magnitude is $25$ mag.
A predicted position of the baryon peak at $z=1(3)$ (see Sec. \ref{sec1}) is $102.1(102.5) \Mpc$, its width $10.2(9.5) \Mpc$ 
and the matter correlation function  amplitude at the peak scale $6.33(3.16) \times 10^{-4}$. 
Then from eq.(\ref{dr1}) we obtain that  the magnitude calibration error $\delta m$ should be smaller than $0.14(0.05)
\times b^2$ for survey of red (blue) galaxies at $z \simeq 1$ if we require an error on the baryon peak position be smaller than $1\%$. 
Note that the calibration depends on the galaxy bias $b$.
Respective calibration error for $z \simeq 3$ yields  $0.01 \times b^2$. 
If we extrapolate results of \cite{2006astro.ph..5302P}
to the bias of Luminous Red Galaxies to $z \simeq 1$ we could expect the bias of this type of galaxies is $b \simeq 2$. 
Blue galaxies are expected to be much less biased \citep{2005ApJ...630....1Z}. 
Thus, the magnitude calibration for the redshift survey of the red galaxy sample at $z \simeq 1$ should not present 
any practical challenge.  The considered effect can have more impact
on the blue-galaxy sample due to the steeper faint end of their 
luminosity function. Also, for a survey at $z \simeq 3$, the required calibration accuracy may be more difficult to 
obtain, although a large bias, of order a few or more \citep{2002ApJ...565...24P,2006ApJ...637..631K}, 
can help. Predictions for $z \simeq 3$ can vary because the limiting
magnitude is on the exponential part of the luminosity function.   

\section{Conclusions}

We have considered three types of possible systematic effects in the estimation of the cosmological 
distances by means of the baryon acoustic peak in the correlation function. Our analysis accounts for 
the nonlinear evolution of structure, the galaxy biasing with respect to the underlying mass and 
the effect of imperfect recovery of the survey selection function. 
In each of these cases we compare systematic to statistical errors that one could expect for 
a $1000$ sq.~degs. and a half of the sky spectroscopic surveys. 

The nonlinear evolution biases the estimated acoustic scale for redshift $z \la 1.5$.  
For redshift $z=1$ (and the depth of the survey $\Delta z = 0.5$)
the systematic error is $0.3\%$ and grows up to $2.4\%$ at $z=0$, close to the predicted 
statistical errors for the hemisphere-scale survey.
Statistical errors are expected to dominate the systematic ones for any survey smaller 
than this, e.g.  the  modest-scale survey yields statistical errors larger by a factor of $4.5$ than 
the hemisphere-scale one. 
Thus, the effect of the nonlinear evolution on the measured acoustic scale is expected to be below
statistical errors for $z \ga 1$ in case of planned surveys even without applying correction 
for nonlinear evolution. Correction techniques, like the one proposed by \citet{2006astro.ph..4361E},
may be useful in unbiasing the acoustic scale even for shallow surveys, $z \ll 1$.   

The measured acoustic scale may also be affected by the scale-dependent galaxy bias which 
is poorly known at present.  We quantified a possible effect of the galaxy bias based
on the toy model supported by the halo model results. We showed that for the case of the galaxy
bias predicted by the halo model it is unlikely to introduce systematic effect on acoustic scale exceeding 
$1\%$ when haloes of mass $\la 3M_{\star}$ at the redshift $z=1$ are considered. 
For less massive haloes the acoustic scale bias is smaller. In fact, the bias for $M_{\star}$ 
haloes at $z=1$ is $0.7\%$ which is lower by a factor of $2$
than statistical errors for the modest-scale survey ($1.5\%$) and higher than those  
expected for the hemisphere-scale survey ($0.3\%$). 
For higher redshifts, the acoustic scale bias for $M_{\star}$ haloes is expected to be at least 
a factor of $2.5(4.75)$ smaller than the statistical errors for the redshift $z=2(3)$ in case 
of the modest-scale survey. 
On the other hand, the systematic error could be larger if the change of the bias with 
scale was more rapid or took place at larger scales than suggested by the halo model. 

Also, we considered the bias in the acoustic scale measurement due to the 
correlated errors in the survey window function. In our analysis the 
systematic error depends linearly upon the local slope of the correlation function of 
the selection function residuals and varying inversely with the local curvature of the true 
correlation function.    
Our analysis implies  that  in order to achieve a bias $<1\%$ 
one has to measure the window function (more precisely, $\xi_{W}(\sim 100 \Mpc)$)
with accuracy better than $2\%$ at $z=1$,  
and more accurately for higher redshifts. The constraint
can be relaxed if the galaxy bias on scales of the acoustic peak is greater than 1.
Specific results depend on details of a survey and a data reduction methods. 
We considered one model, namely independent photometric zero-point errors at each 
telescope pointing. RMS zero-point errors (at limiting magnitude $25$ mag) below  
$0.14$ mag and $0.01$ mag for redshift $z=1$ (red galaxies) and $z=3$ (Lyman-break galaxies), 
respectively, are required in order to limit systematic error in the peak-of-the-correlation-function 
estimator to less than $1\%$. 
In order to gain a confidence in unbiasedness of the acoustic scale measurement,
techniques to diagnose window function problems have to be developed and applied
\citep{2002MNRAS.335..887T}.  

We compared the variance of the peak-of-the-correlation function estimator 
to the power spectrum estimator \citep{2006MNRAS.365..255B} and found the former
yields virtually the same statistical inaccuracies as the latter. 
This suggests that our model-independent estimator might be useful in measuring of 
the sound horizon scale in spite of the simplifying assumptions we made in the course 
of its analysis. Even though its variance may turn out to be larger,
its simplicity and model independence are sufficient reasons for the 
peak-of-the-correlation method to be practically important.  

In this paper we considered only the real space correlation function. 
In the redshift space, a peak in the correlation function becomes  
an elliptical ridge and one has to properly modify the peak-of-the-correlation function 
estimator to become applicable in this case.  
Fortunately, the large scale redshift space distortions \citep{1987MNRAS.227....1K} 
do not lead to a bias of the baryon-peak position because the relation between an azimuthally averaged
redshift space power spectrum and the real space one is scale independent. However,  
an improved analysis of redshift space distortions 
\citep{2001MNRAS.325.1359S, 2004PhRvD..70h3007S} may lead to different implications. 
We defer an analysis of the redshift space effects to the future work.

\section*{Acknowledgments}
We would like to thank Bhuvnesh Jain, Mike Jarvis, Hee-Jong Seo and Ravi Sheth  
for numerous helpful discussions.
This work is supported by NSF grant AST-0236702, Department of Energy grant
DOE-DE-FG02-95ER40893, NASA grant BEFS-04-0014-0018  and 
Polish State Committee for Scientific Research grant 1P03D01226.


\begin{thebibliography}{}

\bibitem[\protect\citeauthoryear{{Angulo}, {Baugh}, {Frenk}, {Bower}, {Jenkins}
  \& {Morris}}{{Angulo} et~al.}{2005}]{2005MNRAS.362L..25A}
{Angulo} R.,  {Baugh} C.~M.,  {Frenk} C.~S.,  {Bower} R.~G.,  {Jenkins} A.,
  {Morris} S.~L.,  2005, \mnras, 362, L25

\bibitem[\protect\citeauthoryear{{Berlind} \& {Weinberg}}{{Berlind} \&
  {Weinberg}}{2002}]{2002ApJ...575..587B}
{Berlind} A.~A.,  {Weinberg} D.~H.,  2002, \apj, 575, 587

\bibitem[\protect\citeauthoryear{{Bernstein}}{{Bernstein}}{2006}]{2006ApJ...63%
7..598B}
{Bernstein} G.,  2006, \apj, 637, 598

\bibitem[\protect\citeauthoryear{{Blake} \& {Glazebrook}}{{Blake} \&
  {Glazebrook}}{2003}]{2003ApJ...594..665B}
{Blake} C.,  {Glazebrook} K.,  2003, \apj, 594, 665

\bibitem[\protect\citeauthoryear{{Blake}, {Parkinson}, {Bassett}, {Glazebrook},
  {Kunz} \& {Nichol}}{{Blake} et~al.}{2006}]{2006MNRAS.365..255B}
{Blake} C.,  {Parkinson} D.,  {Bassett} B.,  {Glazebrook} K.,  {Kunz} M.,
  {Nichol} R.~C.,  2006, \mnras, 365, 255

\bibitem[\protect\citeauthoryear{{Cole} et~al.}{{Cole} et~al.}{2005}]{2005MNRAS.362..505C}
{Cole} S. et~al.,  2005, \mnras, 362, 505

\bibitem[\protect\citeauthoryear{{Cooray} \& {Sheth}}{{Cooray} \&
  {Sheth}}{2002}]{2002PhR...372....1C}
{Cooray} A.,  {Sheth} R.,  2002, \physrep, 372, 1

\bibitem[\protect\citeauthoryear{{Eisenstein} \& {Hu}}{{Eisenstein} \&
  {Hu}}{1998}]{1998ApJ...496..605E}
{Eisenstein} D.~J.,  {Hu} W.,  1998, \apj, 496, 605

\bibitem[\protect\citeauthoryear{{Eisenstein}, {Seo} \& {White}}{{Eisenstein}
  et~al.}{2006}]{2006astro.ph..4361E}
{Eisenstein} D.~J.,  {Seo} H.-j.,    {White} M.,  2006, astro-ph/0604361

\bibitem[\protect\citeauthoryear{{Eisenstein} et~al.}{{Eisenstein} et~al.}{2005}]{2005ApJ...633..560E}
{Eisenstein} D.~J. et~al.,  2005, \apj, 633, 560

\bibitem[\protect\citeauthoryear{{Faber} et~al.}{{Faber} et~al.}{2005}]{2005astro.ph..6044F}
{Faber} S.~M. et~al., 2005, astro-ph/0506044

\bibitem[\protect\citeauthoryear{{Feldman}, {Kaiser} \& {Peacock}}{{Feldman}
  et~al.}{1994}]{1994ApJ...426...23F}
{Feldman} H.~A.,  {Kaiser} N.,    {Peacock} J.~A.,  1994, \apj, 426, 23

\bibitem[\protect\citeauthoryear{{Guzik} \& {Bernstein}}{{Guzik} \&
  {Bernstein}}{2005}]{2005PhRvD..72d3503G}
{Guzik} J.,  {Bernstein} G.,  2005, \prd, 72, 043503

\bibitem[\protect\citeauthoryear{{Hamilton}}{{Hamilton}}{2000}]{2000MNRAS.312.%
.257H}
{Hamilton} A.~J.~S.,  2000, \mnras, 312, 257

\bibitem[\protect\citeauthoryear{{Hu} \& {Jain}}{{Hu} \&
  {Jain}}{2004}]{2004PhRvD..70d3009H}
{Hu} W.,  {Jain} B.,  2004, \prd, 70, 043009

\bibitem[\protect\citeauthoryear{{Kaiser}}{{Kaiser}}{1987}]{1987MNRAS.227....1%
K}
{Kaiser} N.,  1987, \mnras, 227, 1

\bibitem[\protect\citeauthoryear{{Kashikawa} et~al.}{{Kashikawa} et~al.}{2006}]
{2006ApJ...637..631K} {Kashikawa} N. et~al., 2006, \apj, 637, 631

\bibitem[\protect\citeauthoryear{{Kravtsov}, {Berlind}, {Wechsler}, {Klypin},
  {Gottl{\"o}ber}, {Allgood} \& {Primack}}{{Kravtsov}
  et~al.}{2004}]{2004ApJ...609...35K}
{Kravtsov} A.~V.,  {Berlind} A.~A.,  {Wechsler} R.~H.,  {Klypin} A.~A.,
  {Gottl{\"o}ber} S.,  {Allgood} B.,    {Primack} J.~R.,  2004, \apj, 609, 35

\bibitem[\protect\citeauthoryear{{Marinoni} et~al.}{{Marinoni} et~al.}{2005}]{2005A&A...442..801M}
{Marinoni} C.  et~al., 2005, \aap, 442, 801

\bibitem[\protect\citeauthoryear{{Matsubara}}{{Matsubara}}{2004}]{2004ApJ...61%
5..573M}
{Matsubara} T.,  2004, \apj, 615, 573

\bibitem[\protect\citeauthoryear{{McDonald}, {Trac} \& {Contaldi}}{{McDonald}
  et~al.}{2005}]{2005astro.ph..5565M}
{McDonald} P.,  {Trac} H.,    {Contaldi} C.,  2005, astro-ph/0505565

\bibitem[\protect\citeauthoryear{{Meiksin}, {White} \& {Peacock}}{{Meiksin}
  et~al.}{1999}]{1999MNRAS.304..851M}
{Meiksin} A.,  {White} M.,    {Peacock} J.~A.,  1999, \mnras, 304, 851

\bibitem[\protect\citeauthoryear{{Miller}, {Nichol} \& {Chen}}{{Miller}
  et~al.}{2002}]{2002ApJ...579..483M}
{Miller} C.~J.,  {Nichol} R.~C.,    {Chen} X.,  2002, \apj, 579, 483

\bibitem[\protect\citeauthoryear{{Padmanabhan} et~al.}{{Padmanabhan} et~al.}%
{2006}]{2006astro.ph..5302P} {Padmanabhan} N. et~al., 2006, astro-ph/0605302

\bibitem[\protect\citeauthoryear{{Porciani} \& {Giavalisco}}{{Porciani} \&
  {Giavalisco}}{2002}]{2002ApJ...565...24P}
{Porciani} C.,  {Giavalisco} M.,  2002, \apj, 565, 24

\bibitem[\protect\citeauthoryear{{Schlegel}, {Finkbeiner} \&
  {Davis}}{{Schlegel} et~al.}{1998}]{1998ApJ...500..525S}
{Schlegel} D.~J.,  {Finkbeiner} D.~P.,    {Davis} M.,  1998, \apj, 500, 525

\bibitem[\protect\citeauthoryear{{Schulz} \& {White}}{{Schulz} \&
  {White}}{2006}]{2006APh....25..172S}
{Schulz} A.~E.,  {White} M.,  2006, Astroparticle Physics, 25, 172

\bibitem[\protect\citeauthoryear{{Scoccimarro}}{{Scoccimarro}}{2004}]{2004PhRv%
D..70h3007S}
{Scoccimarro} R.,  2004, \prd, 70, 083007

\bibitem[\protect\citeauthoryear{{Seljak}}{{Seljak}}{2000}]{2000MNRAS.318..203%
S}
{Seljak} U.,  2000, \mnras, 318, 203

\bibitem[\protect\citeauthoryear{{Seljak}}{{Seljak}}{2001}]{2001MNRAS.325.1359%
S}
{Seljak} U.,  2001, \mnras, 325, 1359

\bibitem[\protect\citeauthoryear{{Seo} \& {Eisenstein}}{{Seo} \&
  {Eisenstein}}{2003}]{2003ApJ...598..720S}
{Seo} H.-J.,  {Eisenstein} D.~J.,  2003, \apj, 598, 720

\bibitem[\protect\citeauthoryear{{Seo} \& {Eisenstein}}{{Seo} \&
  {Eisenstein}}{2005}]{2005ApJ...633..575S}
{Seo} H.-J.,  {Eisenstein} D.~J.,  2005, \apj, 633, 575

\bibitem[\protect\citeauthoryear{{Sheth}, {Hui}, {Diaferio} \&
  {Scoccimarro}}{{Sheth} et~al.}{2001}]{2001MNRAS.325.1288S}
{Sheth} R.~K.,  {Hui} L.,  {Diaferio} A.,    {Scoccimarro} R.,  2001, \mnras,
  325, 1288

\bibitem[\protect\citeauthoryear{{Smith} et al.}{{Smith} et al.}{2003}]
{2003MNRAS.341.1311S}
{Smith} R.~E. et al., 2003, \mnras, 341, 1311

\bibitem[\protect\citeauthoryear{{Smith}, {Scoccimarro} \& {Sheth}}{{Smith}
  et~al.}{2006}]{2006astro.ph..9547S}
{Smith} R.~E.,  {Scoccimarro} R.,    {Sheth} R.~K.,  2006, astro-ph/0609547

\bibitem[\protect\citeauthoryear{{Spergel}, {Verde}, {Peiris}, {Komatsu},
  {Nolta}, {Bennett}, {Halpern}, {Hinshaw}, {Jarosik}, {Kogut}, {Limon},
  {Meyer}, {Page}, {Tucker}, {Weiland}, {Wollack} \& {Wright}}{{Spergel}
  et~al.}{2003}]{2003ApJS..148..175S}
{Spergel} D.~N. et al., 2003, \apjs, 148, 175

\bibitem[\protect\citeauthoryear{{Springel}, {White}, {Jenkins}, {Frenk},
  {Yoshida}, {Gao}, {Navarro}, {Thacker}, {Croton}, {Helly}, {Peacock}, {Cole},
  {Thomas}, {Couchman}, {Evrard}, {Colberg} \& {Pearce}}{{Springel}
  et~al.}{2005}]{2005Natur.435..629S}
{Springel} V. et al., 2005, \nat, 435, 629

\bibitem[\protect\citeauthoryear{{Steidel}, {Adelberger}, {Giavalisco},
  {Dickinson} \& {Pettini}}{{Steidel} et~al.}{1999}]{1999ApJ...519....1S}
{Steidel} C.~C.,  {Adelberger} K.~L.,  {Giavalisco} M.,  {Dickinson} M.,
  {Pettini} M.,  1999, \apj, 519, 1

\bibitem[\protect\citeauthoryear{{Tegmark}}{{Tegmark}}{1997}]{1997PhRvL..79.38%
06T}
{Tegmark} M.,  1997, Physical Review Letters, 79, 3806

\bibitem[\protect\citeauthoryear{{Tegmark} et~al.}{{Tegmark}
  et~al.}{2006}]{2006astro.ph..8632T}
{Tegmark} M. et~al., 2006, astro-ph/0608632

\bibitem[\protect\citeauthoryear{{Tegmark}, {Hamilton} \& {Xu}}{{Tegmark}
  et~al.}{2002}]{2002MNRAS.335..887T}
{Tegmark} M.,  {Hamilton} A.~J.~S.,    {Xu} Y.,  2002, \mnras, 335, 887

\bibitem[\protect\citeauthoryear{{Tinker}, {Weinberg}, {Zheng} \&
  {Zehavi}}{{Tinker} et~al.}{2005}]{2005ApJ...631...41T}
{Tinker} J.~L.,  {Weinberg} D.~H.,  {Zheng} Z.,    {Zehavi} I.,  2005, \apj,
  631, 41

\bibitem[\protect\citeauthoryear{{Yang}, {Mo} \& {van den Bosch}}{{Yang}
  et~al.}{2003}]{2003MNRAS.339.1057Y}
{Yang} X.,  {Mo} H.~J.,    {van den Bosch} F.~C.,  2003, \mnras, 339, 1057

\bibitem[\protect\citeauthoryear{{Zehavi}  et~al.}{{Zehavi}  et~al.}{2005}]{2005ApJ...630....1Z}
{Zehavi} I. et~al., 2005, \apj, 630, 1

\bibitem[\protect\citeauthoryear{{Zheng}}{{Zheng}}{2004}]{2004ApJ...610...61Z}
{Zheng} Z., 2004, \apj, 610, 61

\bibitem[\protect\citeauthoryear{{Zheng} et al.}{{Zheng} et al.}{2005}
]{2005ApJ...633..791Z}
{Zheng} Z. et al.,  2005, \apj, 633, 791

\end{thebibliography}

\end{document}